\documentclass[aps,prl,twocolumn,superscriptaddress,floatfix]{revtex4-1}
\usepackage{amsmath}
\usepackage{amssymb}
\usepackage{amsthm}
\usepackage{amsfonts}
\usepackage{listings}
\usepackage{enumerate}
\usepackage{latexsym}
\usepackage{psfrag}
\usepackage{bm}
\usepackage[all]{xy}
\usepackage{graphicx}
\usepackage{subfigure}
\usepackage[pdftex,colorlinks=false]{hyperref}
\usepackage{color}
\usepackage{mathtools}
\usepackage{eufrak}
\usepackage{ gensymb }
\usepackage{lipsum}
\begin{document}
	
	\title{From Claringbullite to a new spin liquid candidate Cu$_3$Zn(OH)$_6$FCl}
	
	\author{Zili Feng}
	\thanks{These two authors contribute to this work equally.}
	\affiliation{Beijing National Laboratory of Condensed Matter Physics and Institute of Physics, Chinese Academy of Sciences, Beijing 100190, China}
	\affiliation{School of Physical Sciences, University of Chinese Academy of Sciences, Beijing 100190, China}	
	
	\author{Wei Yi}
	\thanks{These two authors contribute to this work equally.}
	\affiliation{Nano Electronics Device Materials group, National Institure for Materials Science (NIMS), 1-1 Namiki, Tsukuba, Ibaraki 305-0044, Japan}
	
	\author{Kejia Zhu}
	\affiliation{Beijing National Laboratory of Condensed Matter Physics and Institute of Physics,
		Chinese Academy of Sciences, Beijing 100190, China}
	
	\author{Yuan Wei}
	\affiliation{Beijing National Laboratory of Condensed Matter Physics and Institute of Physics,
		Chinese Academy of Sciences, Beijing 100190, China}
	\affiliation{School of Physical Sciences, University of Chinese Academy of Sciences, Beijing 100190, China}
	
	\author{Shanshan Miao}
	\affiliation{Beijing National Laboratory of Condensed Matter Physics and Institute of Physics,	Chinese Academy of Sciences, Beijing 100190, China}
	
	\author{Jie Ma}
	\affiliation{Key Laboratory of Artificial Structures and Quantum Control, School of Physics and Astronomy, Shanghai Jiao Tong University, Shanghai 200240, China}
	\affiliation{Collaborative Innovation Center of Advanced Microstructures, Nanjing, Jiangsu 210093, China}
	
	\author{Jianlin Luo}
	\affiliation{Beijing National Laboratory of Condensed Matter Physics and Institute of Physics,
		Chinese Academy of Sciences, Beijing 100190, China}
	\affiliation{School of Physical Sciences, University of Chinese Academy of Sciences, Beijing 100190, China}
	\affiliation{Collaborative Innovation Center of Quantum Matter, Beijing 100190, China}
	
	\author{Shiliang Li}
	\email{slli@iphy.ac.cn}
	\affiliation{Beijing National Laboratory of Condensed Matter Physics and Institute of Physics, Chinese Academy of Sciences, Beijing 100190, China}
	\affiliation{School of Physical Sciences, University of Chinese Academy of Sciences, Beijing 100190, China}
	\affiliation{Collaborative Innovation Center of Quantum Matter, Beijing 100190, China}
	
	\author{Zi Yang Meng}
	\email{zymeng@iphy.ac.cn}
	\affiliation{Beijing National Laboratory of Condensed Matter Physics and Institute of Physics, Chinese Academy of Sciences, Beijing 100190, China}
	\affiliation{CAS Center of Excellence in Topological Quantum Computation and School of Physical Sciences, University of Chinese Academy of Sciences, Beijing 100190, China}
	\affiliation{Songshan Lake Materials Laboratory, Dongguan, Guangdong 523808, China}
	
	\author{Youguo Shi}
	\email{ygshi@iphy.ac.cn}
	\affiliation{Beijing National Laboratory of Condensed Matter Physics and Institute of Physics,
		Chinese Academy of Sciences, Beijing 100190, China}
	\affiliation{School of Physical Sciences, University of Chinese Academy of Sciences, Beijing 100190, China}
		
	\date{\today}
	
	\begin{abstract}
				The search for quantum spin liquid (QSL) materials has attracted significant attention in the field of condensed matter physics in recent years, but until now only a handful of them are considered as candidates hosting QSL ground state. Owning to their geometrically frustrated structure, Kagome materials are ideal system to realize QSL. In this study, we synthesized the kagome structured material Claringbullite (Cu$_4$(OH)$_6$FCl) and then  replaced inter-layer Cu with Zn to form Cu$_3$Zn(OH)$_6$FCl. Comprehensive measurements reveal that doping Zn$^{2+}$ ions transforms magnetically ordered Cu$_4$(OH)$_6$FCl into a non-magnetic QSL candidate Cu$_3$Zn(OH)$_6$FCl. Therefore, the successful syntheses of Cu$_4$(OH)$_6$FCl and Cu$_3$Zn(OH)$_6$FCl not only provide a new platform for the study of QSL but also a novel pathway of investigating the transition between QSL and magnetically ordered systems.
	\end{abstract}

\date{\today}

\maketitle

In recent years, the search for quantum spin liquid (QSL) materials, usually realized in frustrated magnets, has attracted great interests due to the exotic anyonic excitations therein as well as their potential relationship with quantum computation and unconventional superconductivity~\cite{Anderson1987,Kitaev2003,Kitaev2006,Balents2010,Savary2016,Zhou2017}.  The common search ground for QSL are frustrated magnets with honeycomb, triangular and kagome structures. Kagome Heisenberg antiferromagnet is a promising direction for the pursuit of QSL ground state. Although several kagome materials are proposed to host QSL ground states, the detailed nature of these QSLs candidates still need further intensive investigations to verify~\cite{Han2012,Fu2015,Norman2016,Feng2017,Wei2017,Feng2018}. 

Among the discovered kagome Heisenberg antiferromagnetic QSL candidates~\cite{Han2011,Feng2017,Zheng2017}, Herbertsmithite, ZnCu$_3$(OH)$_6$Cl$_2$~\cite{Imai2008,Imai2011,Han2012,Norman2016,Han2016} and Zn-doped Barlowite, Cu$_3$Zn(OH)$_6$FBr~\cite{Feng2017,WenXG2017,Wei2017,Feng2018,SMAHA2018,Ranjith2018,McQueen2018} are the two well-known kagome QSL materials.  In particular, the high-quality single crystal of Herbertsmithite provides the rare opportunity to explicitly reveal the detailed momentum-frequency dependence of the magnetic spectra in inelastic neutron scattering experiment, in which the spinon continuum manifest~\cite{Han2012}. Meanwhile, many related materials are also studied, such as Cu$_3$Mg(OH)$_6$Cl$_2$, Cu$_3$Cd(OH)$_6$SO$_4$, Y$_3$Cu$_9$(OH)$_{19}$Cl$_8$ and YCu$_3$(OH)$_6$Cl$_3$ ~\cite{Nocera2005,Wills2011,Thomale2015,Pustogow2017,Krellner2017,Sun2016}. To synthesize a promising kagome QSL material, the perfect kagome structure must retain intact and the inter-layer interaction shall be reduced, to optimize the perfect frustration in the 2D plane in order to suppress any magnetic order. Unfortunately, most materials develop magnetic order at low temperatures which is usually accompanied by structure distortion that breaks the perfect kagome lattice geometry. 

In this study, we successfully synthesized a new kagome Heisenberg antiferromagnetic QSL candidate, Cu$_3$Zn(OH)$_6$FCl and its parent compound Cu$_4$(OH)$_6$FCl (Claringbullite). X-ray diffraction (XRD) shows pure Cu$_3$Zn(OH)$_6$FCl is synthesized and chemical analysis show that the content of  Zn$^{2+}$ ions in the inter-kagome plane is about 0.7 per formula unit. Although some Cu$^{2+}$ ions remain in the inter-kagome plane, structure analysis reveals that the perfect kagome plane is preserved and thermodynamic measurements show that magnetic order is completely suppressed in Cu$_3$Zn(OH)$_6$FCl. Moreover, our AC magnetic susceptibility (ACMS) measurements also exclude spin glass behavior in Cu$_3$Zn(OH)$_6$FCl, which is found in other putative QSL candidate~\cite{JSWen2018}. All these evidence suggests Cu$_3$Zn(OH)$_6$FCl is a new QSL candidate. On the other hand, Claringbullite develops magnetic order below 15 K. The evolution from Cu$_4$(OH)$_6$FCl to Cu$_3$Zn(OH)$_6$FCl, therefore, provides an ideal occasion to investigate the transition between magnetically ordered states and QSL, in which the theoretically expected dynamical signatures of fractionalized anyonic excitations~\cite{Mei2015,Sun2018,Becker2018} and other exotic properties of the associated Z$_2$ topologically ordered QSLs~\cite{WenZ2SL1991,Wen2017} could be experimentally revealed in future.

Cu$_4$(OH)$_6$FCl and Cu$_3$Zn(OH)$_6$FCl were synthesized using hydrothermal method. For Cu$_4$(OH)$_6$FCl, about 0.55 g Cu$_2$(OH)$_2$CO$_3$ (Alfa Aeser), 0.43 g CuCl$_2\cdot$2H$_2$O (Alfa Aeser), 0.185 g NH$_4$F (Alfa Aeser) and 0.1 g concentrated hydrochloric acid (HCl, mass fraction about 35$\%$) were mixed with about 20 mL deionized water in a 25 mL vessel.  For Cu$_3$Zn(OH)$_6$FCl, about 0.5g Cu$_2$(OH)$_2$CO$_3$, 0.38 g ZnCl$_2\cdot$4H$_2$O (Alfa Aeser), 0.095g NH$_4$F and 0.1 g HCl were also mixed with about 20 mL deionized water in a 25 mL vessel. The samples both were heated at 200 $^\circ C$ for one day and then slowly cooled to room temperature. Green powder ware collected by washing and drying the production.

\label{tab:table1}
\begin{table}
	\begin{tabular*}{0.9\columnwidth}{cccccccc}
	\hline
		Site & $w$ & $x$ & $y$ & $z$&$B$  \\ 
		\hline
		Cu & $6g$ &0.5&0&0&1.794(31)  \\
		Zn & $2d$ &1/3 &2/3 &3/4 &2.290(117) \\
		Cl & $2c$ &2/3 &1/3 &3/4 &1.336(112)\\
		F  & $2b$ &0.0 &0.0 &3/4 &1.767(125)\\
		O  & $12k$ &0.20076&0.79924(24) &0.90986(27)&0.112(89)\\
		H  & $12k$ &0.12472&0.87528 &0.86683 &2.265\\
		\hline	
	\end{tabular*}
	\caption{Structure parameters of Cu$_3$Zn(OH)$_6$FCl at room temperature. The space group is P6$_3$/mmc (No. 194). Z = 2, a = b = 6.65377(11) \AA, c = 9.18896(11) \AA, and V = 352.3166 \AA$^{3}$. The occupancy of Zn site is 0.319 and the others are 1 according to a splitting model~\cite{Han2014, He2018}. The R indexes are $R_{wp}$ = 4.583 \% and $R_p$ = 3.259 \%.}
\end{table}

\begin{figure}
	\centering \includegraphics[width=\columnwidth]{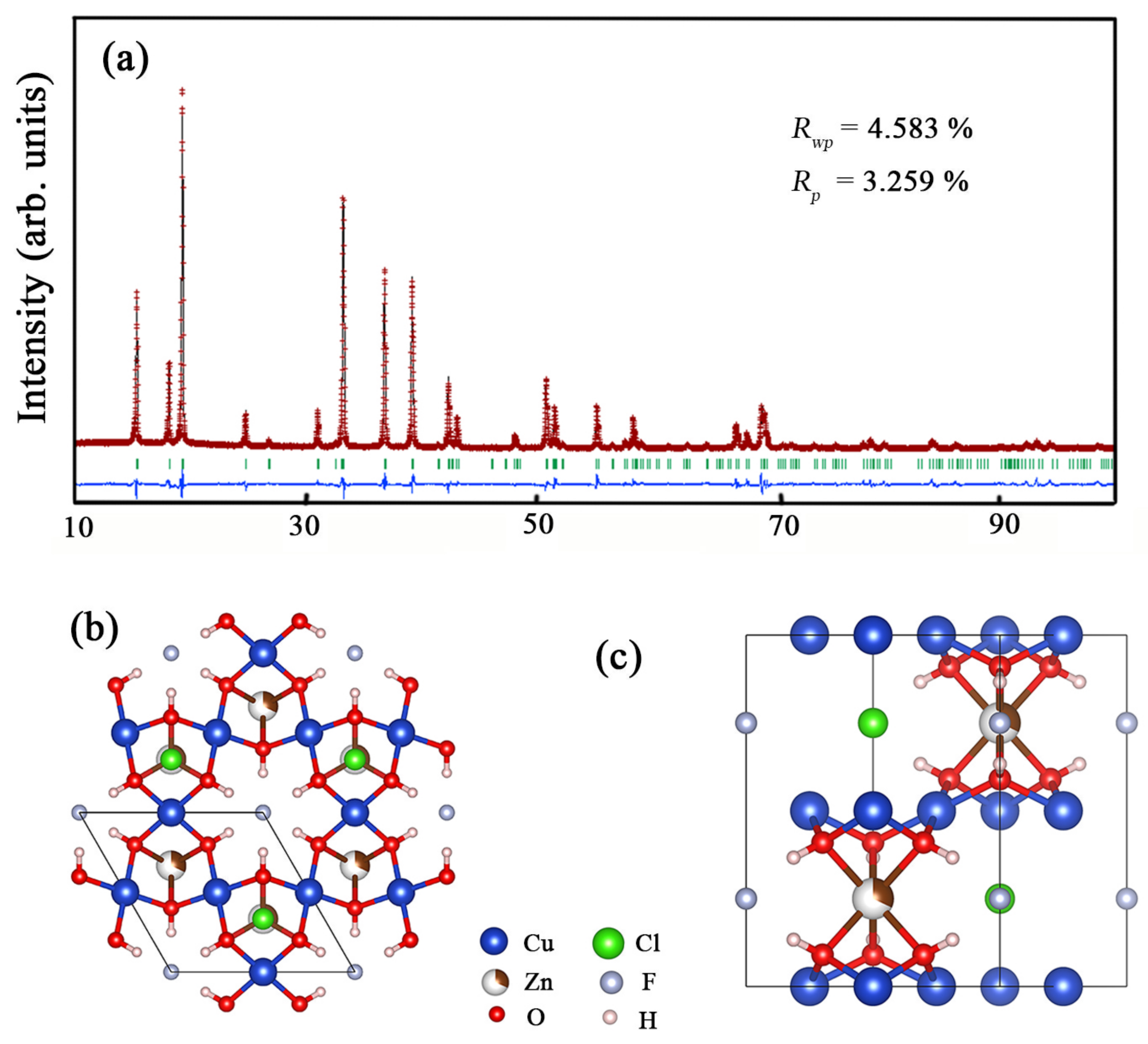}
	\caption{XRD pattern and structure of Cu$_3$Zn(OH)$_6$FCl. (a) Powder XRD data and Rietveld refinement profile of Cu$_3$Zn(OH)$_6$FCl. (b) Top view (ab plane) of the structure of Cu$_3$Zn(OH)$_6$FCl. (c) Slide view of the structure of Cu$_3$Zn(OH)$_6$FCl. }
	\label{fig:figure1}
\end{figure}

Cu$_3$Zn(OH)$_6$FCl was characterized by powder XRD and no impurity peaks were found. The measured XRD data and  the calculated pattern, refined by the Rietveld method using the program RIETAN-FP ~\cite{Rietveld1969} are plotted in Fig.~\ref{fig:figure1} (a). The refined structure parameters are listed in Table 1. The XRD measurement was performed at room temperature on a Rigaku Smartlab high-resolution diffractometer using Cu K$\alpha_1$ radiation ($\lambda$ = 1.5406 \AA). Two forms of chemical component analysis were used to determine the Zn content. First, energy-dispersive X-ray spectra were measured on a Hitachi S-4800 scanning electron microscope at an accelerating voltage of 15 kV, with an accumulation time of 90 s. Then, inductively coupled plasma atomic emission spectroscopy was performed using a Thermo IRIS Intrepid II instrument. These results show that the stoichiometric ratio of Cu and Zn ions is about 3:0.7.

The sample was pressed and cut into pieces about 1 mm$^2$  to perform the thermodynamic measurements. The magnetization data were measured using a superconducting quantum interference device magnetometer (Quantum Design MPMS). The specific heat data were measured in a physical property measurement system (PPMS) at zero magnetic field between 2 and 30 K and the specific heat below 2 K was measured using a He-3 system.

Cu$_3$Zn(OH)$_6$FCl crystallized in hexagonal lattice with space group P6$_3$/mmc (No. 194). The lattice parameters are a = b = 6.65377(11) \AA, c = 9.18896(11) \AA. As shown in Fig.~\ref{fig:figure1}, Cu$_3$Zn(OH)$_6$FCl has a perfect kagome lattice at room temperature. The kagome planes show AA stacking and are separated by Zn$^{2+}$ ions. Table 1 provides detailed structure parameters of Cu$_3$Zn(OH)$_6$FCl. In Cu$_4$(OH)$_6$FCl, the interlayer Cu$^{2+}$ ions split from the equilibrium site and shows three equivalent positions~\cite{He2018}, while in Cu$_3$Zn(OH)$_6$FCl, the interlayer Zn$^{2+}$ ions sit at the center of the triangle of kagome Cu layer.

\begin{figure}
	\centering	\includegraphics[width=\columnwidth]{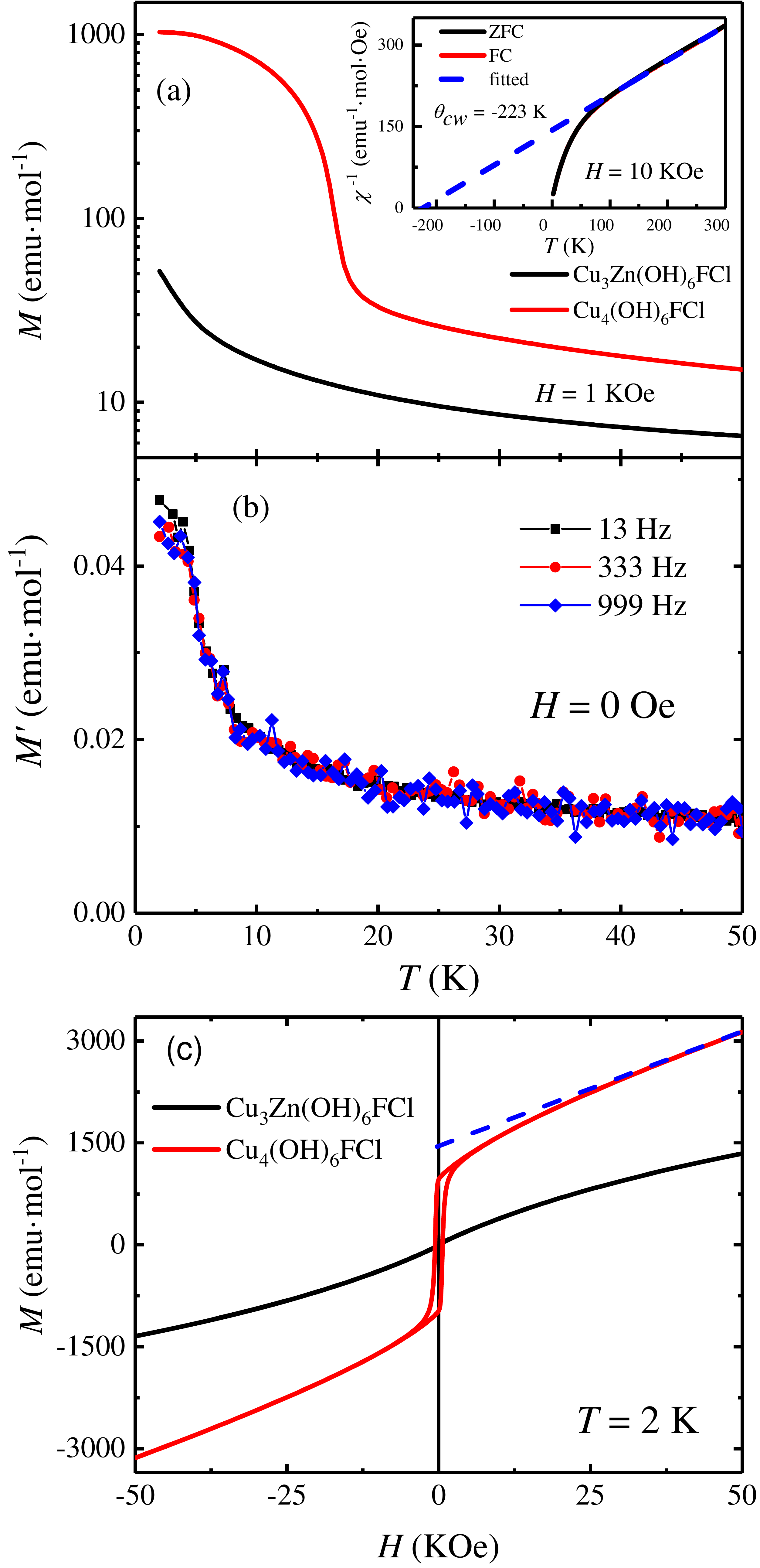}
	\caption{Magnetic properties of Cu$_3$Zn(OH)$_6$FCl and Cu$_4$(OH)$_6$FCl. (a). Temperature dependence of magnetic moment between 2 K and 50 K at 1 KOe for Cu$_3$Zn(OH)$_6$FCl and Cu$_4$(OH)$_6$FCl. Inset: zero-field-cooling and field-cooling $1/\chi$ at 10 KOe for Cu$_3$Zn(OH)$_6$FCl. Curie-Weiss fitting (blue dashed line) gives the Curie temperature $\theta_{CW}$= -223 K. (b). The temperature dependence of the real component of ACMS for Cu$_3$Zn(OH)$_6$FCl. These data were measured under zero magnetic field and the peak amplitude of applied excitation field was 1 Oe. (c). Field dependence of magnetic moment of Cu$_3$Zn(OH)$_6$FCl and Cu$_4$(OH)$_6$FCl at 2 K. The blue dashed line is linear fitting ($M = A+B\times H$) of high field magnetization for Cu$_4$(OH)$_6$FCl.}
	\label{fig:figure2}
\end{figure}

Fig.~\ref{fig:figure2} (a) shows the temperature dependence of magnetic moment of Cu$_3$Zn(OH)$_6$FCl and Cu$_4$(OH)$_6$FCl at 1 KOe below 50 K. The vertical coordination is in logarithmic scale to highlight the salient difference. The moment of Cu$_4$(OH)$_6$FCl increases abruptly at about 17 K, which is its magnetic phase transition temperature as reported in previous study~\cite{He2018}, while Cu$_3$Zn(OH)$_6$FCl shows no obvious magnetic order transition down to 2 K. Furthermore, the ACMS measurements at zero magnetic field show no frequency dependence down to 2 K as shown in Fig.~\ref{fig:figure2} (b), which suggests no spin-glass behavior~\cite{JSWen2018}. The $1/\chi$ data between 300 K and 2 K of Cu$_3$Zn(OH)$_6$FCl at 10 KOe in zero-field-cooling (ZFC) and field-cooling (FC) conditions are shown in the inset of Fig.~\ref{fig:figure2} (a).  Curie-Weiss law ($\chi = c / (T - \theta_{cw}$)) was used to fit the data between 150 K and 300 K and gave $\theta_{cw} = - 223$ K. The value of $\theta_{cw}$ (19.2 meV) indicates a strong antiferromagnetic interaction between Cu$^{2+}$ ions in the kagome layer, suggesting strong frustration in the system. The Curie constant is 1.556 K$\cdot$emu$\cdot$mol$^{-1}$.  Assuming $S = 1/2$ and using the typical $g$ factor value in kagome lattice for Cu ions $g= 2.3\pm0.1$\cite{Han2014,Feng2017}, the Cu content can be calculated, $3.28\pm0.14$ per formula unit. This gives that Zn content is $0.72\pm0.14$ per formula unit which is consistent with the chemical analysis discussed above.

Fig.~\ref{fig:figure2} (c) shows the field dependence of magnetization of Cu$_3$Zn(OH)$_6$FCl and Cu$_4$(OH)$_6$FCl at 2 K. Cu$_4$(OH)$_6$FCl shows obvious ferromagnetic behavior with the magnetization jump about 0.0865 $\mu_B$/Cu, while Cu$_3$Zn(OH)$_6$FCl shows no visible hysteresis.  The comparison between Cu$_4$(OH)$_6$FCl and Cu$_3$Zn(OH)$_6$FCl in Fig.~\ref{fig:figure2} (c) demonstrates the disappearance of magnetic order in the latter, although the Zn content is not equal to 1. A linear function $M = A+B\times H$ is used to fit the magnetization at high field as shown in Fig.~\ref{fig:figure2} (c). Similar fitting analysis has been performed in Cu$_{4-x}$Zn$_x$(OH)$_6$FBr and the Zn content dependence (doping dependence) of $B(x)$ has been obtained in Fig. 6(b) in Ref.~\cite{Feng2018}. Compared with the fitting results in Fig.~\ref{fig:figure2} (c) here, we find very consistent value in the $x$-dependence of the Zn content. For Cu$_4$(OH)$_6$FCl, we obtain $B = 0.035$ emu$\cdot$mol$^{-1}$, almost the same as that in Cu$_4$(OH)$_6$FBr; and for Cu$_3$Zn(OH)$_6$FCl, we obtain $B = 0.02$ emu$\cdot$mol$^{-1}$, this value is very close to the $x=0.7$ value in Cu$_{4-x}$Zn$_x$(OH)$_6$FBr, again suggesting the success of synthesizes of  Cu$_3$Zn(OH)$_6$FCl which may be a new QSL candidate.

\begin{figure}
	\centering	\includegraphics[width=\columnwidth]{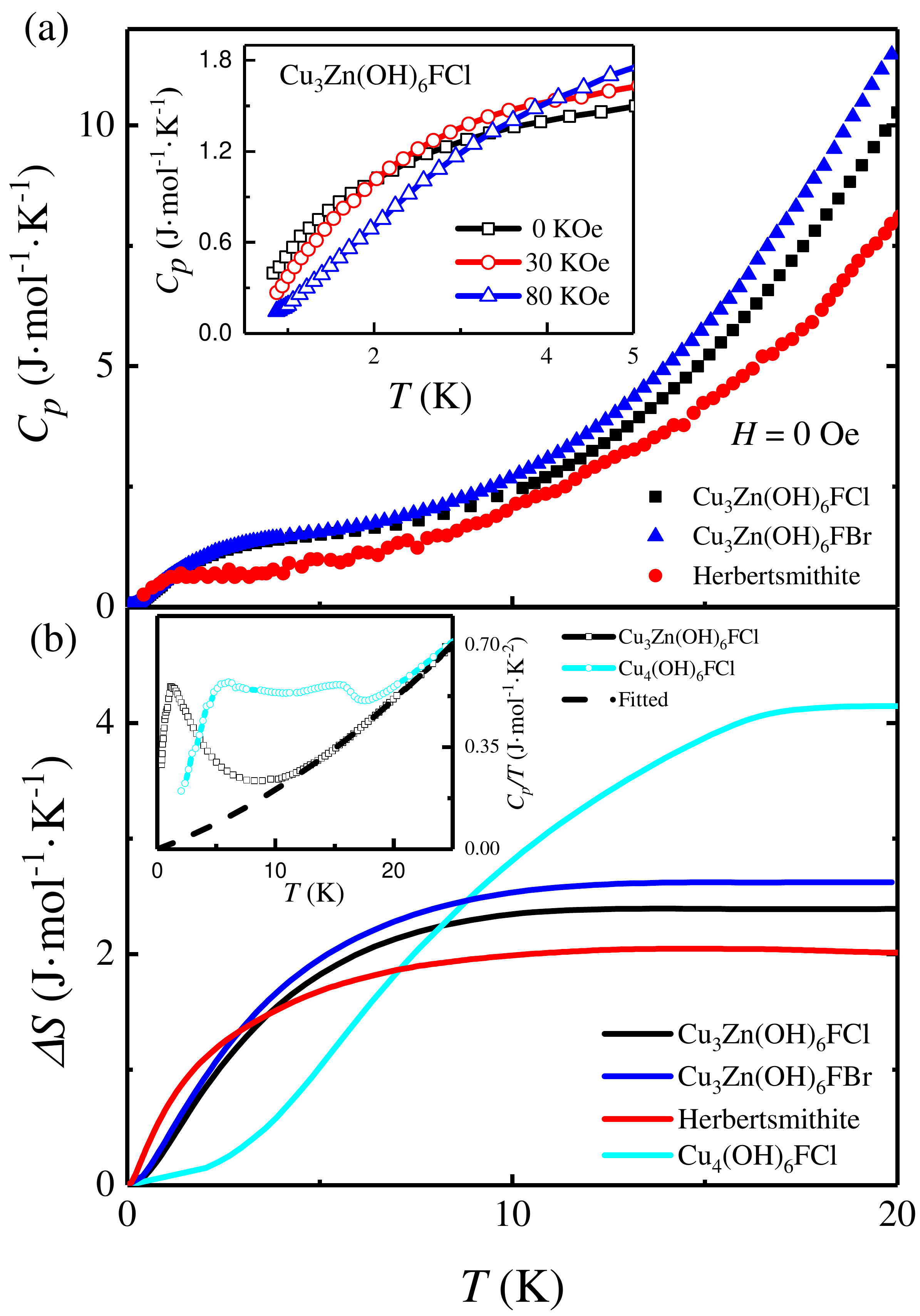}
	\caption{Thermal dynamic properties of Cu$_3$Zn(OH)$_6$FCl, Cu$_3$Zn(OH)$_6$FBr, Herbertsmithite, and Cu$_4$(OH)$_6$FCl. (a). Specific heat at zero magnetic field over temperature for Cu$_3$Zn(OH)$_6$FBr (blue), Cu$_3$Zn(OH)$_6$FCl (black), and Herbertsmithite (red). The data of Cu$_3$Zn(OH)$_6$FBr was from Fig.2(c) in ref.\cite{Feng2017} and that of Herbertsmithite was from Fig.3(a) in ref.\cite{Han2014}. The inset plots the low temperatures behavior of Cu$_3$Zn(OH)$_6$FCl at 0 Oe, 30 KOe and  80 KOe. (b) Magnetic entropy of Cu$_3$Zn(OH)$_6$FCl, Cu$_3$Zn(OH)$_6$FBr, Herbertsmithite and Cu$_4$(OH)$_6$FCl. The inset shows the pecific heat over temperature at zero field below 25 K of Cu$_3$Zn(OH)$_6$FCl and Cu$_4$(OH)$_6$FCl. The dashed line are background contribution of Cu$_3$Zn(OH)$_6$FCl fitted using $C_{bg} = a T^2+b T^3$.}
	\label{fig:figure3}
\end{figure}

Similar contrasts have also been reported in details in the comparisons in the Zn doped Cu$_{4-x}$Zn$_x$(OH)$_6$Cl$_2$ which shows hysteresis and QSL candidate Herbertsmithite Cu$_3$Zn(OH)$_6$Cl$_2$~\cite{deVries2008}, as well as the more recent Zn doped Barlowite Cu$_{4-x}$Zn$_x$(OH)$_6$FBr and the QSL candidate Cu$_3$Zn(OH)$_6$FBr~\cite{Feng2018}. Such consistency further reaffirms that all these three kagome material families, Cu$_{4-x}$Zn$_x$(OH)$_6$Cl$_2$, Cu$_{4-x}$Zn$_x$(OH)$_6$FBr and Cu$_{4-x}$Zn(OH)$_{x}$FCl, are ideal platforms of achieving kagome QSL with $x\to1$. Moreover, our results are consistent with a previous study on Cu$_4$(OH)$_6$FCl, in which a magnetic transition at around 15 K was reported~\cite{He2018}. Recent neutron scattering and NMR studies reveal that Cu$_4$(OH)$_6$FBr change its structure at low temperature~\cite{Feng2018,Ranjith2018} and such structural change gives rise to magnetic order. It is possible that a similar process happened in Cu$_4$(OH)$_6$FCl as well. Further neutron scattering experiment, similar to that performed on Cu$_4$(OH)$_6$FBr~\cite{Wei2017,Feng2018}, will reveal this process. 

Fig.~\ref{fig:figure3} (a) shows the specific heat of Cu$_3$Zn(OH)$_6$FCl at zero field from 0.8 K to 20 K. For comparison, the data of Cu$_3$Zn(OH)$_6$FBr (from Fig.2(c) in Ref.~\cite{Feng2017}) and Herbertsmithite (from Fig.3(a) in Ref.~\cite{Han2014}) were also plotted. These three compounds shows similar behavior at low temperatures. Especially, Cu$_3$Zn(OH)$_6$FCl and Cu$_3$Zn(OH)$_6$FBr are nearly coincide below 10 K.  The inset shows the specific heat of Cu$_3$Zn(OH)$_6$FCl below 5 K at different fields and no magnetic order was observed. As for Cu$_4$(OH)$_6$FCl, the magnetic order transition arises at about $15 \sim 17$ K, which is manifested by the high-temperature peak of $C_{p}/T$ showed in the inset of Fig.~\ref{fig:figure3} (b). Further reduce the temperature, $C_{p}/T$ of Cu$_4$(OH)$_6$FCl shows another peak at 5 K, and $C_{p}/T$ of Cu$_3$Zn(OH)$_6$FCl shows a peak at around 3 K. These low-temperatur peaks, have also been seen in Cu$_{4-x}$Zn$_x$(OH)$_6$FBr~\cite{Feng2018,SMAHA2018} and previous study of Cu$_4$(OH)$_6$FCl~\cite{He2018}, as well as in Herbertsmithite~\cite{Helton2007,deVries2008,Han2011}. They are all significantly affected by magnetic field, as exemplified in the inset of Fig.~\ref{fig:figure3} (a) for Cu$_3$Zn(OH)$_6$FCl and the inset of Fig.2 (c) for Cu$_3$Zn(OH)$_6$FBr in Ref.~\cite{Feng2017}. The current understanding of such behavior is that these are the contribution from inter kagome layer Cu ions~\cite{Han2016,Feng2017,Feng2018}, which have very weak interactions among themselves and are easily polarized by external field. 
 
 One can obtained some information of kagome QSL from the current data by substrate the background contribution using the formula $C_{bg} = a T^2+b T^3$. This formula was used to fit the specific heat data between 20 K and 30 K in zero field for Cu$_3$Zn(OH)$_6$FCl, Cu$_4$(OH)$_6$FBr, Herbertsmithite, and Cu$_4$(OH)$_6$FCl. The dashed line in the inset of Fig.\ref{fig:figure3} (b) shows the fitted data for Cu$_3$Zn(OH)$_6$FCl.   The $T^2$ term comes from the spin correlations above 20 K in two dimensional structure and the $T^3$ term, on the other hand, comes from the three dimensional lattice contribution~\cite{Han2014}. Fig.\ref{fig:figure3} (b) shows magnetic entropy of Cu$_3$Zn(OH)$_6$FCl, Cu$_3$Zn(OH)$_6$FBr, Herbertsmithite and Cu$_4$(OH)$_6$FCl. At about 20 K, their magnetic entropies reach corresponding maximal values. The entropy of Cu$_3$Zn(OH)$_6$FCl lays between the values of Cu$_3$Zn(OH)$_6$FBr and Herbertsmithite,  while for Cu$_4$(OH)$_6$FCl, its magnetic entropy is very close to that of Cu$_4$(OH)$_6$FBr~\cite{Feng2018}. These results indicate that Cu$_3$Zn(OH)$_6$FCl is a QSL candidate. 

To summarize, in this work we have successfully synthesized a new QSL candidate material Cu$_3$Zn(OH)$_6$FCl and its parent material Claringbullite Cu$_4$(OH)$_6$FCl and performed comprehensive thermodynamic measurements on them. We find Cu$_3$Zn(OH)$_6$FCl is a highly frustrated kagome material and do not show magnetic order down to 0.8 K and hence is a good candidate for kagome QSL, whereas Cu$_4$(OH)$_6$FCl is magnetically ordered below 17 K. The possible structural change in Cu$_4$(OH)$_6$FCl accompanied with the development of magnetic order is worth to be revealed by elastic neutron scattering. Our results consistently reveal the similarity of the three by now established families of the kagome Heisenberg antiferromagnets, Cu$_{4-x}$Zn$_x$(OH)$_6$Cl$_2$, Cu$_{4-x}$Zn$_x$(OH)$_6$FBr and Cu$_{4-x}$Zn(OH)$_{x}$FCl. As $x\to1$, Heisenberg kagome QSL properties of Cu$_3$Zn(OH)$_6$FCl emerge similar to those of Herbertsmithite Cu$_3$Zn(OH)$_6$Cl$_2$~\cite{Han2011} and Zn-doped Barlowite Cu$_3$Zn(OH)$_6$FBr~\cite{Feng2017}.

Looking into the future, the pathway from Cu$_4$(OH)$_6$FCl to Cu$_3$Zn(OH)$_6$FCl offers the opportunity to investigate the transition between magnetically ordered systems to QSL state. Such investigations will have a very significant theoretical impact. As the material realization of Z$_2$ topological order, neutron scattering experiments on the possible kagome QSL in Cu$_3$Zn(OH)$_6$FCl and its transition to the magnetically order Claringbullite Cu$_4$(OH)$_6$FCl, could help to reveal the dynamical signature of the fractionalized anyonic excitations in the QSL ground state, as proposed by theoretical calculations at vicinity of the QSL to magnetic order phase transition~\cite{Sun2018,Becker2018}. In this regard, the materials such as Cu$_4$(OH)$_6$FCl and Cu$_3$Zn(OH)$_6$FCl presented in this work, will greatly encourage the further theoretical and experimental developments of the new paradigms of quantum matter.

This research was supported by the Ministry of Science and Technology of China through the National Key
Research and Development Program (2016YFA0300502, 2017YFA0302901, 2016YFA0300604 and 2016YFA0300501), the Strategic Priority Research Program of the Chinese Academy of Sciences (XDB28000000, XDB07020100 and QYZDB-SSW-SLH043), the National Natural Science Foundation of China (No.11421092, 11574359, 11674370, 11774399, 11474330 and U1732154).

\bibliography{Cu3Zn(OH)6FCl}

\clearpage
\end{document}